\documentstyle[a4wide,12pt,epsf]{article}

\title{Extended Variational Approach to the $SU(2)$ Mass Gap 
on the Lattice.}

\author{\\
J.~O.~Akeyo \\ 
Physics Department\\
Kenyatta University \\ 
P.O.~Box 43844 \\ 
Nairobi, Kenya \\
\\
H.~F.~Jones and C.~S.~Parker \\
Physics Department,\\
Imperial College,\\
London SW7 2BZ, UK\\
}

\date {}
\begin{document}    
\bibstyle{unrst}
\maketitle

\begin{abstract}
The linear delta expansion is applied to a calculation of
the $SU(2)$ mass gap on the lattice.  Our results compare
favourably with the strong-coupling expansion and are in
good agreement with recent Monte Carlo estimates.
\end{abstract}
    
\def\thepage{Imperial/TP/93-94/37}
				   
\thispagestyle{myheadings}

\newpage
\advance\topmargin -0.5in 
\baselineskip=24pt 
\pagenumbering{arabic}
\addtocounter{page}{1}

\subsection*{1. Introduction}  

The linear delta expansion (LDE) is an analytic approach to field
theory which has been applied to a number of different problems (see
for example Ref. \cite{Jones_Conf}).  The approach is
non-perturbative in the sense that a power series expansion is made
in a parameter $\delta$  artificially inserted into the action,
rather than in a coupling constant of the theory.  The calculational
techniques required do not differ greatly from conventional Feynman
diagrams.  An essential part of the approach is an optimization with
respect to another parameter, in the present case $J$, appearing in 
the $\delta$-extended action.

The linear delta expansion uses $\delta$ as an interpolation between a
soluble action $S_0$ and the action for the desired theory $S$.
The action is written: 
\begin{equation}
    S_{\delta}  =  (1 - \delta ) S_{0}  + \delta S,
    \label{eq-lde_def}
\end{equation} 
where $S_0$ contains some dependence on the optimization parameter
$J$.  A vacuum generating functional or appropriate Green function may
then be evaluated as a power series in $\delta$, which is set equal to
unity at the end of the calculation.

Of course this power series is only calculated to a finite number of
terms, and will therefore retain some dependence on $J$ which would be 
absent in the sum to all orders when $\delta$ is set equal to one.
 A well-motivated criterion for fixing 
$J$ is to demand that, at least locally, the truncated result should be 
independent of $J$. This is the principle of minimal sensitivity (PMS)
\cite{PMS}. If $C_N$ denotes the $N$th approximant to a quantity $C$, 
the requirement is 
\begin{equation}     
    \frac{\partial C_N (J)}{\partial J} = 0     
    \label{eq-pms}   
\end{equation}

This, or some similar criterion, is an intrinsic part of the LDE, 
providing 
the non-perturbative dependence on the coupling constant of the 
theory.  For example, in the delta expansion of the integral 
$\int dx \exp (-g x^4) $, the PMS correctly reproduces its $g^{-1/4}$
dependence. The application of the PMS is also vital for the convergence 
of the $\delta$ series, which has been rigorously proved for the 
zero-dimensional $\phi^4$ vacuum generating functional \cite{Buc_Dunc_Jones} 
and the finite temperature partition function of the anharmonic oscillator 
in quantum mechanics \cite{Dunc_Jones}. The proof has been recently 
extended \cite{Ben_Dunc_Jones} to the connected
vacuum generating function $W = \ln Z$ in zero dimensions.

A number of non-perturbative approaches to field theory are related to
the LDE.  At first order in $\delta$, the LDE is related to the
Gaussian approximation \cite{Gauss}, and at higher orders to
generalizations of this \cite{Oko}. It also has much in common with
work of Kleinert \cite{HK} and of Sissakian et al. \cite{SSS}.

In the context of lattice gauge theories the LDE has been applied, with
various choices of the trial action \cite{lde_GN}-\cite{Buc_Jones_2}, 
to the groups $Z_2$, $U(1)$ and $SU(2)$, 
mainly in calculating the plaquette energy $E_P$.  A particularly useful
trial action is the one proposed by  Zheng et al.~\cite{Zheng}, 
based on single links:
\begin{equation}
    S_{0} = J \sum _{\ell} {\rm tr} U_{\ell}
    \label{eq-trial_links} 
\end{equation} 
The sum runs over all links $\ell$ of the lattice, and the parameter
$J$ is used for optimization. However, these authors used a different
optimization criterion, more closely related to the conventional
variational method, in which a rigorous inequality for the free
energy at $O(\delta)$ was applied at {\em all orders in} $\delta$.
Such a procedure is liable to forfeit the convergence which may be
provided by an order-by-order optimization. 

Two of the present authors \cite{Ak_Jones_1} used the Zheng trial
action with the PMS in its usual sense in a
calculation of the $SU(2)$ plaquette energy.  This was found to give 
excellent agreement
at $O(\delta^3)$ with Monte Carlo results in the
weak coupling regime.  
Following on from this, the phase structure of the
mixed $SU(2)$ - $SO(3)$ action was studied \cite{Ak_Jones_2}, and again
the results to $O(\delta^3)$ gave good agreement with the Monte Carlo
results.  

We were therefore encouraged to attempt to extend the method to the more 
difficult problem of the mass gap in lattice $SU(2)$ using the same 
trial action. For such quantities, which involve finding the exponential 
fall-off of a correlator at large separations,  semi-analytic methods 
have, in principle, an advantage over Monte-Carlo methods, insofar as the
size of the lattice is not limited and small signals are not masked by
statistics. 

 In Section 2 we set up the formalism for the problem to be 
studied and explain how the diagrams which arise in the delta expansion 
of the modified action are evaluated. The optimization procedure adopted
is explained in Section 3, where the results are presented first in lattice 
units, and then in terms of the SU(2) lattice constant $\Lambda_L$ by 
looking for
the correct scaling limit as $a\rightarrow 0$. In Section 4 we
summarize the paper and indicate some directions for further development. 
The appendix shows how the evaluation of expectation values in the
background of the trial action of Eq.~(\ref{eq-trial_links}) can be
organized in a way amenable to symbolic computation.

\newpage
\subsection*{2. The Diagrammatic Expansion}

We consider the $SU(2)$ gauge theory on the lattice.  The $\delta$-extended
action is :
\begin{equation}
\label{eq-action} 
S_\delta = \frac{1}{2} \delta\beta \sum_P {\rm tr} \, U_P +(1-\delta)
		J \sum_\ell {\rm tr} \, U_\ell
\end{equation}
The partition function for this system may be written:
\begin{eqnarray}
\label{eq-parfn}
	Z_\delta & = & \int [dU] \, {\rm e}^{S_\delta} \nonumber \\ 
	& = & \int [dU] \, \sum _{r=0} ^\infty
	\delta^r\frac{(S-S_0)^r}{r!} {\rm e}^{S_0}
\end{eqnarray} 
and lattice quantities may be evaluated as power series in $\delta$
in the background $S_0$.  This leads to a diagrammatic expansion
related, but not identical to the conventional strong coupling
$\beta$ expansion  \cite{Creutz} \cite{Muenster}, the difference being
that the strong coupling expectations are evaluated in a zero
background.  The actual diagrams used are also different, the first
non-vanishing diagram in the strong coupling expansion for the mass gap
being a closed cuboid of plaquettes, compared with the LDE, for which the
first diagram is shown in Fig.~1(a).

Calculation of the mass gap involves the evaluation of the connected
correlation $C(t)$ between two non-oriented plaquettes $A$ and $B$ 
with temporal separation $ta$ in any spatial position.
\begin{equation}
\label{eq-corr}
	C(t) = < {\rm tr} U_B(ta) {\rm tr} U_A(0) >_C 
\end{equation} 
The subscript $C$ denotes the connected expectation or cumulant.  
The diagrammatic
expansion in powers of $\delta$ has its first non-vanishing term at
$O(\delta^t)$.  This is shown in
Fig.~1(a), where a ``ladder" of time-like plaquettes connects $A$ and $B$.

The next power in $\delta$ adds one extra plaquette to
Fig.~1(a) in all possible positions.  Some examples
of these are shown as Figs.~1(b)-(k).  It should be
noted that this calculation is carried out in the temporal gauge. 
This explains the absence of diagrams where a plaquette is attached to
the side of the ladder by a temporal link only. Such a link variable is set
to unity, and therefore the extra plaquette is effectively disconnected.

At this order, there is also a term proportional to $S^t S_0$ in the 
$\delta$ expansion.  This can be included as a partial derivative with 
respect to $J$ of the $O(\delta^t)$ diagram.

Each of the diagrams shown has an associated multiplicity depending
on its geometric properties. The basic diagram of
Fig.~1(a) has a factor of $4R_1$ representing the fact that the ladder
may be connected to any of the four sides of the lower plaquette $A$,
which is taken as fixed, and that the upper plaquette $B$ has $R_1$
possible spatial orientations. Similarly the additional factor of
$12R_2$ in Fig.~1(b) arises from the $R_2$ possible spatial
orientations of the extra plaquette, its possible attachment on any of
the three sides of its neighbour, the fact that either of the two
upper plaquettes could be $B$, and finally a factor of two to include
the symmetrical configuration where the extra plaquette is attached to
$A$ instead. Note that Figs.~1(f), (h) and (k),
which involve an additional plaquette in the body of the ladder, have 
a $t$-dependent multiplicity.  

Having enumerated the diagrams to the required order and calculated
their associated multiplicities, their
expectation values must be evaluated.  The evaluation of simpler diagrams
consisting of up to four or five plaquettes by group integration
\cite{Zheng} or character expansion \cite{Ak_Jones_2} has been
discussed elsewhere.  Another method is discussed in the appendix to
the present paper.  

In the evaluation of the straight ladder diagram of Fig.~1(a) and
subsequent modifications thereof, an enormous simplification arises
from the fact that the expectation value of a single link is a
multiple of the identity (Eq.~(A1)). This means that the contribution
of Fig.~1(a) is just a product of factors representing the expectation
values of the doubled links occurring on each rung. Remarkably this
factorization extends to the connected expectation value, with the
result that
\begin{equation}
\label{C1a}
C_{\rm 1(a)} = (4R_1)\;2V_2^6V^{t+1}\delta^t
\end{equation}
Here each factor of $V_2$ represents the expectation value of a single
spacelike link of $A$ and $B$, and we get a factor of $V\equiv
\partial V_2/\partial J$ for each rung. The functions $V_n$ are
defined as ratios of modified Bessel functions of argument $2J$:

\begin{equation}
	V_n (J)  =  \frac{I_n(2J)}{I_1(2J)} 
        \label{eq-Vdefs}
\end{equation}

Two derived quantities which appear frequently in the contributions of
higher order diagrams are $V$ as defined above:
\begin{equation}
	V (J)  =  \frac{\partial V_2}{\partial J}
\end{equation}
and
\begin{equation}
	Y (J)  =  4 V_4 - 6 V_3 + 2 V_2
        \label{eq-VY}
\end{equation}

The higher order diagrams consist of modifications to this basic
diagram by inserting additional plaquettes at either end and/or in the
middle. At order $\delta^{t+1}$ only one of these alternatives is
possible. The factorization property noted above extends to these
higher order diagrams. That is, their connected expectation can be
obtained from the basic building blocks shown in Fig.~2, with
additional factors representing the bulk of the ladder.

Diagrams which involve a modification at one end have a multiplicity
which is independent of the total temporal separation $t$, whereas the
three diagrams 1(f), 1(h) and 1(k) which involve an addition to the
middle of the ladder have $t$-dependent multiplicities. The latter
essentially exponentiate in higher orders and so are the only ones
which contribute to the mass gap when this is calculated from the
$\delta$-expansion of the ratio $C(t+1)/C(t)$ (See Eq.~(\ref{eq-tayrat})).
The contributions of these diagrams are given below:
\begin{eqnarray}
\label{order t+1}
C_{\rm 1(f)} & = & (12R_1^2 (t-1))\;V^{t+2}V_2^8 \delta^{t+1}\nonumber \\
C_{\rm 1(h)} & = & (4R_1^2 (t-1))\;V^tV_2^9(Y-8V_2V) \delta^{t+1} \\
C_{\rm 1(k)} & = & (2R_1 t)\;V^{t-1}V_2^6(Y^2/3-16V_2^2V^2)\delta^{t+1}
\nonumber
\end{eqnarray}

The diagrams of order $\delta^{t+2}$ are similarly built up by adding a
further plaquette in all possible ways. Some examples are shown in
Fig.~3. There are around 150 diagrams at this order, although again it
is only those with $t$-dependent multiplicities which contribute to
the $\delta$-expansion of $C(t+1)/C(t)$. There are also additional terms
in $S^t S_0^2$ and $S^{t+1} S_0$ which arise from the expansion of the 
factor $(S-S_0)^r$ in Eq.~\ref{eq-parfn}.
The most succinct way of including such contributions
is to note that $J$ always occurs in the combination $J(1-\delta )$
(See Eq.~(\ref{eq-action})). Thus the $V$'s occurring in the various 
expectation values are really functions of this argument, which needs 
to be Taylor expanded to the appropriate order.
Altogether we may write
\begin{equation}
C(t)=\sum_i M_i D_i(J(1-\delta))\label{order t+2}
\end{equation}
where $M_i$ denotes the multiplicity of the $i$th diagram and $D_i$
is its connected expectation value.

\newpage
\subsection*{3. Extraction of the mass gap}

Having set up the diagrams necessary to calculate the correlation
$C(t)$, we now need to extract the mass gap $m$ using the
familiar result:
\begin{equation}
    C(t) \sim A\;{\rm e}^{-mat}
    \label{eq-mgdef1}
\end{equation}
as $t\rightarrow \infty$, giving
\begin{equation}
    ma = \lim _{t\rightarrow \infty} \ln \frac {C(t)}{C(t+1)}
    \label{eq-mgdef2} 
\end{equation}
 
At first sight it might seem reasonable to calculate $C(t)$ and $C(t+1)$
separately, applying the PMS to each correlation, and then to extract $ma$ from
equation (\ref{eq-mgdef2}). However, this is not a fruitful procedure 
for two reasons.  The first is that it is not in the spirit of the
PMS, according to which it is the final quantity calculated which
should be optimized with respect to  $J$. More importantly, the
convergence of the expansion performed in this manner is extremely slow. 
It is, after all, asking a great deal of a perturbation expansion, even 
when optimized,
to give the correct $t\rightarrow\infty$ limit of $C(t)$ with only a
few terms of the expansion.

The most important aspect of the problem is that some of the diagrams have
multiplicities which grow with $t$, reflecting the fact that additional
plaquettes can be attached in a large number of positions to the body of the
ladder. Thus the larger the value of $t$, the higher the order of the 
perturbation expansion required before the factorial denominators in 
Eq.~(\ref{eq-parfn}) eventually control the convergence. However, these
diagrams essentially exponentiate. For example the series of ``bracket''
diagrams starting with Fig.~1(h) and continuing with Fig.~3(h) has the 
form of an exponential series for large $t$. Consequently, when the series
for the {\it ratio} is taken the $t$-dependence cancels, as we show in more 
detail below. Thus by considering the Taylor expansion for the ratio, the
limit $t\rightarrow\infty$ does not pose such a threat to the convergence
of the series. A similar procedure was adopted by M{\"u}nster \cite{Muenster}
in the application of the strong coupling expansion to the calculation of
the mass gap.

We therefore apply the PMS to the Taylor expansion of the ratio 
$C(t+1) / C(t)$, up to third order in
$\delta$. 
Writing the series for $C(t)$ and $C(t+1)$ as
\begin{eqnarray}
    C(t) &=& \delta^t (b_0 + b_1 \delta + b_2 \delta^2 + \dots )\nonumber \\
    C(t+1)&=& \delta^{t+1} (a_0 + a_1 \delta + a_2 \delta^2 + \dots ), 
    \label{eq-taylor}
\end{eqnarray}
the ratio has the expansion 
\begin{equation}
    \frac{C(t+1)}{C(t)}  = 
    \frac{a_0}{b_0}\delta + \frac {a_1 b_0 - a_0
    b_1}{b_0^2}\delta^2 +\frac{b_1(a_0 b_1-a_1 b_0) -b_0(a_0 b_2 -
    a_2 b_0)}{b_0^3} \delta^3 + \dots  
    \label{eq-tayrat}
\end{equation}

This formulation leads to a na{\"\i}ve large $t$ limit for the mass gap.
In going from temporal separation $t$ to $t+1$, we add an extra
plaquette to the ladder part of each diagram, which gives an overall 
extra factor $\frac{1}{2}\beta V(J)$ to the correlation.
Thus one might expect the mass gap to be equal to $-\ln (\frac{1}{2}
\beta V(J))$.  In fact, for the lowest-order contribution we have 
\begin{eqnarray}
	a_0 & = & f_0 \lambda ^{t+1} \nonumber \\
	b_0 & = & f_0 \lambda ^t     
\end{eqnarray}
where $\lambda = {1 \over 2}\beta V$, so that indeed $a_0/b_0 = 
{1 \over 2}\beta V$.  
In higher orders, however, $t$-dependent multiplicities give rise to
corrections to the na{\"\i}ve result.

The next-order coefficients have the form:
\begin{eqnarray}
	a_1 &=& (f_1 + (t+1)g_1)\lambda^{t+1} \nonumber \\
	b_1 &=& (f_1 + tg_1)\lambda^t 
	\label{eq-tplus1} 
\end{eqnarray}
Then to second order in $\delta$, Eq.~(\ref{eq-tayrat}) is:
\begin{equation}
	\frac{C(t+1)}{C(t)}  =  \delta [ \lambda + \delta
	\frac{\lambda g_1}{f_0}] 	
\end{equation}
Again this is independent of $t$, and means that in this form of the 
expansion $t$ does not need to be taken asymptotically large. It is sufficient to take it large enough for the diagrams to settle down to a generic form.

At $O(\delta^{t+2})$, writing
\begin{eqnarray}
	a_2 &=& (f_2 +(t+1)g_2 + (t+1)^2 h_2) \lambda^{t+1} \nonumber \\
	b_2 &=& (f_2 +tg_2 + t^2 h_2) \lambda^t\;, 
	\label{eq-tplus2}
\end{eqnarray}
the $O(\delta^3)$ term in Eq.~(\ref{eq-tayrat}) is:

\begin{equation}
	\frac{b_1(a_0 b_1-a_1 b_0) -b_0(a_0 b_2 - a_2 b_0)}{b_0^3}
	= \frac{\lambda}{f_0^2} [ f_0(h_2+g_2) - f_1g_1 + t(2h_2f_0 -
	g_1^2)] 	
	\label{eq-d2term}
\end{equation}

This apparently has a $t$-dependence, but in fact the coefficient $h_2$ is 
precisely ${1 \over 2}g_1^2/f_0$ because it arises from exponentiation of 
the $t$-dependent graphs at order $\delta^{t+1}$. As emphasized by M\"unster 
\cite{Muenster}, the summation over the spatial positions of the upper 
plaquette $B$, which also serves to project out zero spatial momentum in the 
correlator, is vital to this exponentiation.

Altogether, then, we have the $t$-independent result for the ratio to order 
$\delta^3$:
\begin{equation}
{C(t+1)\over C(t)} = \delta\lambda \left[ 1 + \delta{g_1 \over f_0}
+ {\delta^2 \over f_0^2}(f_0(h_2+g_2) - f_1g_1)\right].
\end{equation}

This expression for $C(t+1)/C(t)$ is still a function of $J$.
According to the PMS criterion, we are looking for stationary points in $J$. 
Typical curves of the $J$ dependence are shown in Figs.~4 and 5 for 
$O(\delta^{t+1})$ and $O(\delta^{t+2})$ respectively.
At $O(\delta^{t+1})$ there is a single maximum, and at $O(\delta^{t+2})$
the value of $C(t+1)/C(t)$ at the maximum is remarkably close to this, 
even though the position in $J$ is quite different.

At this stage our result is expressed in terms of the inverse of the lattice 
spacing $a$, which we need to take to zero in order to make contact with the 
continuum limit. The physical value of the glueball mass must in this limit
become a fixed number times the SU(2) lattice scale $\Lambda_L$. 
In the weak coupling 
limit, to two-loop level, this is given by  

\begin{equation}
    \Lambda _L a =  \left(\frac{6 \pi^2 \beta}{11} \right)^{51/121} 
                \exp \left( - \frac{3 \pi^2 \beta}{11}\right)
    \label{eq-wcrg}
\end{equation}
We look for the constant $C_m$ such that
\begin{equation}
    ma= C_m \Lambda_L a\;,
    \label{eq-wcrg2}
\end{equation}
which is the value for which the graph of Eq.~(\ref{eq-wcrg}) against 
$\beta$ is
tangential to that of $ma$ calculated in the LDE.  The graphs are shown
in Fig.~6 to $O(\delta^{t+1})$ and Fig.~7 
to $O(\delta^{t+2})$.  These show good agreement
between the orders, the tangents occurring at $\beta = 2.62$ and
$\beta = 2.64$ at $O(\delta^{t+1})$ and $O(\delta^{t+2})$ respectively.  
These results then give for the mass gap:
\begin{eqnarray}
    m & = & 184 \Lambda_L \nonumber \;\;\;\;\;\; (O(\delta^{t+1}))\\
    m & = & 197 \Lambda_L \;\;\;\;\;\; (O(\delta^{t+2}))
    \label{eq-results}
\end{eqnarray}

Compared to the strong-coupling expansion \cite{Muenster}, which gives
$m= 193\Lambda_L$ ($\beta\approx 2.3$) at order $\beta^6$ and 
$m=127\Lambda_L$ ($\beta\approx 1.9$) at order $\beta^8$,
our results show better consistency between consecutive orders; 
moreover, the $\beta$-values where the tangents occur are further into
the weak-coupling region. In a
series expansion of this kind it is difficult to quote a precise error, but
based on the difference between our two results at $O(\delta^{t+1})$ and 
$O(\delta^{t+2})$ one would estimate the error as not more than $\pm 13$.

Our results can be compared directly with those of Berg and Billoire 
 \cite{Berg}, who quote $m = (190 \pm 10)\Lambda_L$.
 A comparison with the more recent work of Michael and Perantonis \cite{MP} 
on a $32^4$ lattice is less straightforward, 
since they quote their results in lattice units and cast some doubt on the
validity of asymptotic scaling. Nonetheless, converting $m=197\Lambda_L$ to
lattice units at $\beta=2.5$ gives $ma=0.70$, in excellent agreement with
their results. At $\beta=2.7$ it gives $ma=0.42$, which is slightly higher
than their central value, but still within the error bars.

\newpage
\subsection*{4. Conclusions}
In this paper we have demonstrated that the linear delta expansion
with the principle of minimal sensitivity is a viable technique for
the calculation of the mass gap for a lattice gauge theory.  We have
shown how the lattice diagrams appearing in this type of calculation
can be easily evaluated by a process of building up chains of
plaquettes from a simple `root' diagram, and that connected
expectations of these are as simple to deal with.  The gauge fixing
procedure adopted reduces the number of contributing 
diagrams, and makes them easier to evaluate. 

As always, the PMS is an integral part of the calculation.
The potential ambiguity arising from the occurrence of multiple PMS points 
is not serious in this case. It is clear by comparison with the lower-order 
calculation that it is the broad maximum at $O(\delta^{t+2})$ which is the 
appropriate one, and it is very encouraging that the resulting value of 
$m$ is so stable in going from one order to the next.

This calculation has shown the relationship between the LDE and the
strong coupling expansion.  The diagrammatic expansion used is
similar, but the actual evaluation of the diagrams is different,
requiring alternative techniques. 

It has proved sufficient to work with the correlators of simple plaquette 
operators rather than the more complicated ``fuzzy'' operators which
have been found necessary in Monte-Carlo calculations. The fundamental 
reason for this is that we are effectively working on an infinite lattice, 
so that large separations are no problem, whereas in Monte-Carlo calculations 
it is necessary to enhance the signal at finite separations. 

The present calculation could be 
extended in various ways. In increasing order of difficulty these are:

\noindent 
(i) to calculate higher mass glueball states. With a simple plaquette
operator the $J^{PC} = 2^{++}$ state occurring in the $E^+$ representation
of the cubic group is 
accessible by weighting the different orientations of the upper plaquette. 
Other spin-parities would require larger Wilson loops.

\noindent
(ii) to work with the gauge group SU(3) rather than SU(2). This would involve
an extension of the techniques of the Appendix to SU(3).

\noindent
(iii) to go to next order in the $\delta$ expansion. The difficulty here
is the greatly increased number of diagrams which have to be taken into
account and the consequent danger of missing an important contribution.

Further possible extensions include calculations of the string tension 
and various quantities at finite temperature. Some work has already been
done on these lines by Tan and Zheng \cite{Zheng2}, but using the free energy 
criterion mentioned above. It would be interesting to return to these problems
 using PMS optimization order by order in the quantity being calculated.

\newpage
\appendix

\section{Appendix}
\label{app}
\renewcommand{\thesection}{\Alph{section}}
\renewcommand{\theequation}{\thesection \arabic{equation}}
\setcounter{equation}{0}

In this appendix we wish to explain a method for calculating
(connected) expectation values of a string of plaquette operators in
the background of our trial action $S_0$ (Eq.~(\ref{eq-trial_links})).  The
method involves expressing the expectation values of products of
single link operators in terms of tensor projection operators
\cite{KM}, which can then be multiplied together within an algebraic
manipulation package.  Of those available, we found FORM the most
suitable because it has an explicit summation convention and extensive
substitution facilities.

Let us start with the simplest example, $<U>$, where $U$ is a
single-link element of $SU(2)$.  It is clear that $<U_a^i> \propto
(Y_{\Box})_a^i := \delta_a^i$, and by taking the trace we establish
the coefficient as $V_2$ (see Eq.~(\ref{eq-Vdefs})):

\begin{equation}
	<U_a^i> = (V_2 Y_{\Box})_a^i
\end{equation}
The notation has been chosen with a view to subsequent examples and in
general refers to the Young tableau associated with a particular
permutation symmetry, in this case trivial, of the upper indices.

For the product of two $U$'s belonging to the same link, there is
similarly no difficulty (again by taking appropriate contractions) in
showing that 
\begin{equation}
	<U_a^i U_b^j>  = (V_3 Y_{\Box\kern-.17em\Box} +
		Y_{\hbox{${\scriptstyle\Box}$\kern-.52em\lower.8ex 
		\hbox{${\scriptstyle\Box}$}}})_{ab}^{ij}
\end{equation}
where
\begin{eqnarray}
	(Y_{\Box\kern-.17em\Box})_{ab}^{ij} = \frac{1}{2}(\delta_a^i \delta_b^j
	+ \delta_a^j \delta_b^i) \nonumber \\
	(Y_{\hbox{${\scriptstyle\Box}$\kern-.52em\lower.8ex 
	\hbox{${\scriptstyle\Box}$}}})_{ab}^{ij} 
	= \frac{1}{2}(\delta_a^i \delta_b^j
	- \delta_a^j \delta_b^i)
\end{eqnarray}

Here $Y_{\Box\kern-.17em\Box}$ and
$Y_{\hbox{${\scriptstyle\Box}$\kern-.52em\lower.8ex
\hbox{${\scriptstyle\Box}$}}}$ are indeed projection operators which
correspond to the two irreducible representations of the permutation
group of the upper indices relative to the lower ones.

If we now go on to the product of three $U$'s, there are three
irreducible representations of the relevant permutation group $S_3$.
However, since we are constructing irreducible tensors of $SU(2)$, the
completely antisymmetric operator 
$Y_{\hbox{${\scriptstyle\Box}$
\kern-.85em\lower.8ex\hbox{${\scriptstyle\Box}$ \kern-.85em\lower.8ex
\hbox{${\scriptstyle\Box}$}}}}$ 
effectively vanishes and only the completely
symmetric and mixed symmetry operators
$Y_{\Box\kern-.17em\Box\kern-.17em\Box}$ and
$Y_{\hbox{${\scriptstyle\Box}$\kern-.17em\hbox{${\scriptstyle\Box}$
\kern-1.2em\lower.8ex\hbox{${\scriptstyle\Box}$}}}}$ 
\, survive.

The form of these tensors can be deduced from the group algebra of
the conjugacy classes of $S_3$, which comprise $E$, $B := 3C_2$ and
$C:=2C_3$.  The non-trivial products in the algebra are $B^2 =
3(E+C)$, $BC = CB = 2B$ and $C^2 = 2E +C$.  The vanishing of 
$Y_{\hbox{${\scriptstyle\Box}$
\kern-.85em\lower.8ex\hbox{${\scriptstyle\Box}$ \kern-.85em\lower.8ex
\hbox{${\scriptstyle\Box}$}}}}$
for $SU(2)$ corresponds to the equivalence relation $B \equiv E+C$ when
acting on the identity element  $\delta_a^i \delta_b^j \delta_c^k$. 
Thus in our search for projection operators we can limit ourselves
to the sub-algebra of $A_3$ generated by the even permutations of
$E$, $C$.  It is then easy to construct the required projection
operators as
\begin{eqnarray}
	Y_{\Box\kern-.17em\Box\kern-.17em\Box} & = & \frac{1}{3} (E+C)
	\nonumber \\
	Y_{\hbox{${\scriptstyle\Box}$\kern-.17em
	\hbox{${\scriptstyle\Box}$\kern-.86em\lower.8ex
	\hbox{${\scriptstyle\Box}$}}}} & = & \frac{1}{3} (2E-C)
\end{eqnarray}
acting on $\delta_a^i \delta_b^j \delta_c^k$ by permutation of the
upper indices.  By taking traces we can establish the coefficients as
$V_4$, $V_2$ in the expansion
\begin{equation}
    < U_a^i U_b^j U_c^k > = (V_4 Y_ {\Box\kern-.17em\Box\kern-.17em\Box}
	+ V_2 
	Y_{\hbox{${\scriptstyle\Box}$\kern-.17em
	\hbox{${\scriptstyle\Box}$\kern-.86em\lower.8ex
	\hbox{${\scriptstyle\Box}$}}}}\,\,)\, 
    	\delta_a^i \delta_b^j \delta_c^k
\end{equation}

For the product of four $U$'s, we can anticipate that
\begin{equation}
    	< U_a^i U_b^j U_c^k U_d^l > = (V_5
	Y_{\Box\kern-.17em\Box\kern-.17em\Box\kern-.17em \Box} + V_3
	Y_{\hbox{${\scriptstyle\Box}$\kern-.17em
	\hbox{${\scriptstyle\Box }$\kern-.17em 
	\hbox{${\scriptstyle\Box}$\kern-1.22em\lower.8ex
	\hbox{${\scriptstyle\Box}$}}}}}\,\,\,\, 
	+ Y_{\hbox{${\scriptstyle\Box
	}$\kern-.17em\hbox{${\scriptstyle\Box}$ 
	\kern-1.2em\lower.8ex\hbox{${\scriptstyle\Box}$\kern-.17em
	\hbox{${\scriptstyle\Box}$}}}}}) 
    	\delta_a^i \delta_b^j \delta_c^k \delta_d^l
\end{equation}
The problem is to establish the specific form of the three
projection operators.  Again the procedure is to look at the group
algebra of $S_4$, which has five conjugacy classes.  Because of the
equivalence relations arising from the vanishing in $SU(2)$ of
completely antisymmetric combinations involving more than two
indices, we can eliminate the classes of odd permutations and work
with the group algebra of the alternating group $A_4$.  This has
three classes, $E$, $B := 3C_2$, $C := 8C_3$, with algebra $B^2 = 3E
+2B$, $BC = CB = 3B$ and $C^2 = 8E+8B+4C$.  From these it is
possible to construct the projection operators
\begin{eqnarray}
    	Y_{\Box\kern-.17em\Box\kern-.17em\Box\kern-.17em \Box}
		& = & \frac{1}{12} (E+B+C) \nonumber \\
    	Y_{\hbox{${\scriptstyle\Box}$\kern-.17em
	\hbox{${\scriptstyle\Box}$\kern-.17em
	\hbox{${\scriptstyle\Box}$\kern-1.22em\lower.8ex
	\hbox{${\scriptstyle\Box}$}}}}} 
		& = & \frac{1}{4} (3E - B) \nonumber \\
    	Y_{\hbox{${\scriptstyle\Box
	}$\kern-.17em\hbox{${\scriptstyle\Box}$ 
	\kern-1.2em\lower.8ex\hbox{${\scriptstyle\Box}$\kern-.17em
	\hbox{${\scriptstyle\Box}$}}}}}
		& = & \frac{1}{12} (2(E+B)-C)
\end{eqnarray}
These formulae are sufficient to evaluate all the diagrams we
encountered up to $O(\delta^{t+2})$.  Diagrams involving $U^{\dag}$
can be dealt with \cite{KM} by converting $U^{\dag}$ to $U$ according to
\begin{eqnarray}
    (U^{\dag})_a^i & = & \epsilon ^{ib} \epsilon_{aj} U_b^j \nonumber \\
                 & = & (\delta_a^i \delta_j^b - \delta_a^b \delta
                       _j^i) U_b^j   
\end{eqnarray}
In particular,
\begin{equation}
    <U_a^i U^{\dag j}_b> = (\frac{1}{2} (V_3 +1) Y_{\Box\kern-.17em\Box}
	+ \frac{1}{2}(3V_3 -1)Y_{\hbox{${\scriptstyle\Box 
	}$\kern-.52em\lower.8ex\hbox{${\scriptstyle\Box}$}}})_{ab}^{ij}
\end{equation}
A given diagram will consist of a number of plaquettes with certain
links in common.  The procedure is then to write down the general
expression for the corresponding amplitude, identify the shared
links, apply the appropriate substitutions for $<UU>$, $<UU^{\dag}>$,
$<UUU>$ and $<UUUU>$ and then sum over all repeated indices.

In fact we actually want the connected expectation values
(cumulants) $< \, >_C$.  These can be obtained by identifying the
different terms in the expansion of 
$\prod _{i=1} ^N (\partial / \partial x_i) \ln Z$ 
with products of expectation values of the
corresponding partitions of the $N$ plaquettes.  Thus
\begin{equation}
    \frac{\partial}{\partial x} \frac{\partial}{\partial y} \ln Z     
       = \frac{1}{Z} \frac{\partial^2 Z}{\partial x \partial y} 
         -  (\frac{1}{Z} \frac{\partial Z}{\partial x})
            (\frac{1}{Z} \frac{\partial Z}{\partial y}) 
\end{equation}
corresponds to 
\begin{equation}
    <U V>_C = <UV> - <U><V>
\end{equation}
etc.

These identifications can all be implemented within a short FORM program.

\newpage

\begin{figure}
	\epsffile{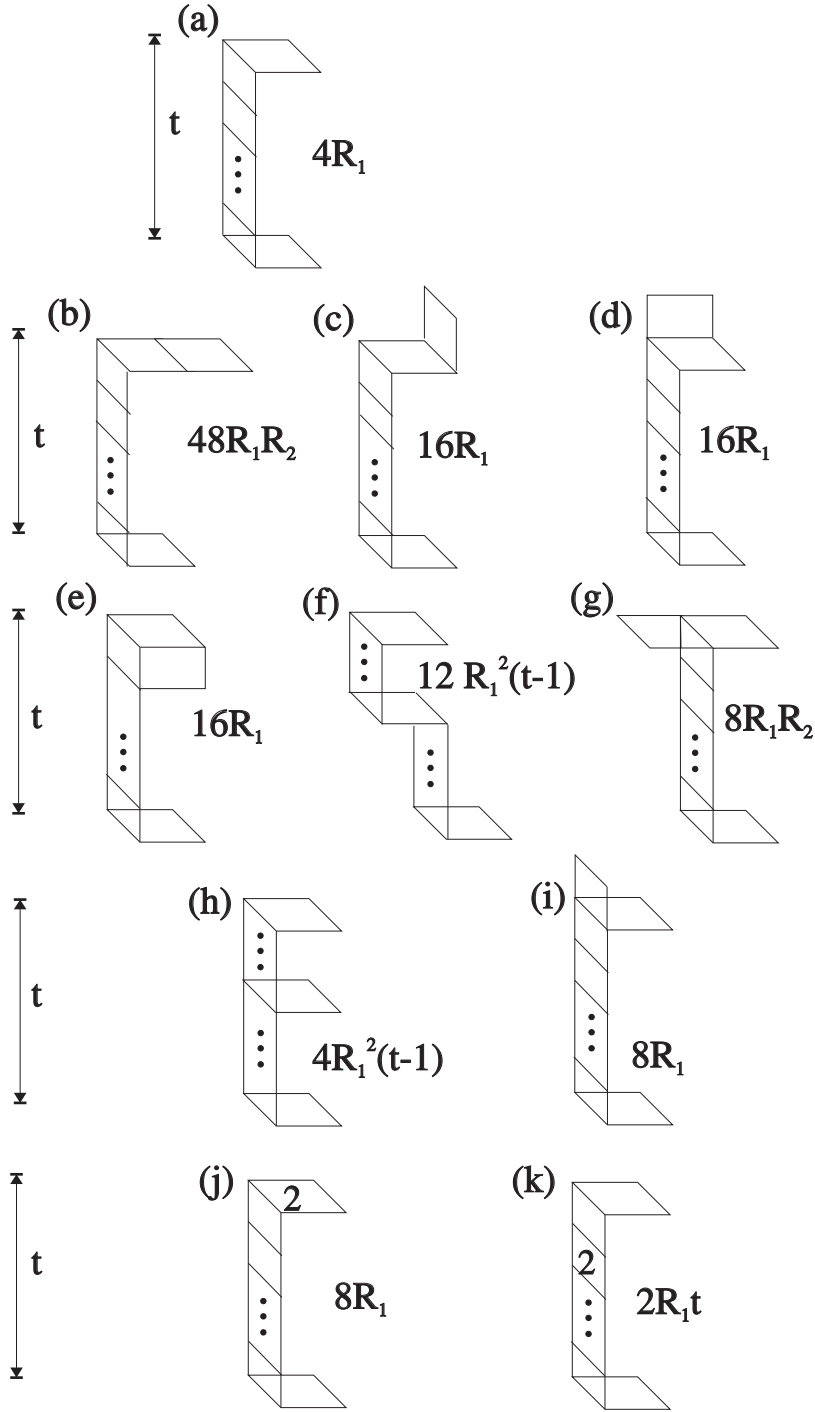} 
        \caption{Non-vanishing diagrams at
	$O(\delta^t)$ and $O(\delta^{t+1})$ in the linear delta
	expansion for the $SU(2)$ mass gap.  The vertical direction
	represents time.  The expression next to each diagram is its
	geometric multiplicity factor.  The $R_i$ are defined by
        $R_i = 2d-3-i$ for $d$ spacetime dimensions.} 
	\label{fig-diags1}
\end{figure}

\begin{figure}
	\epsffile{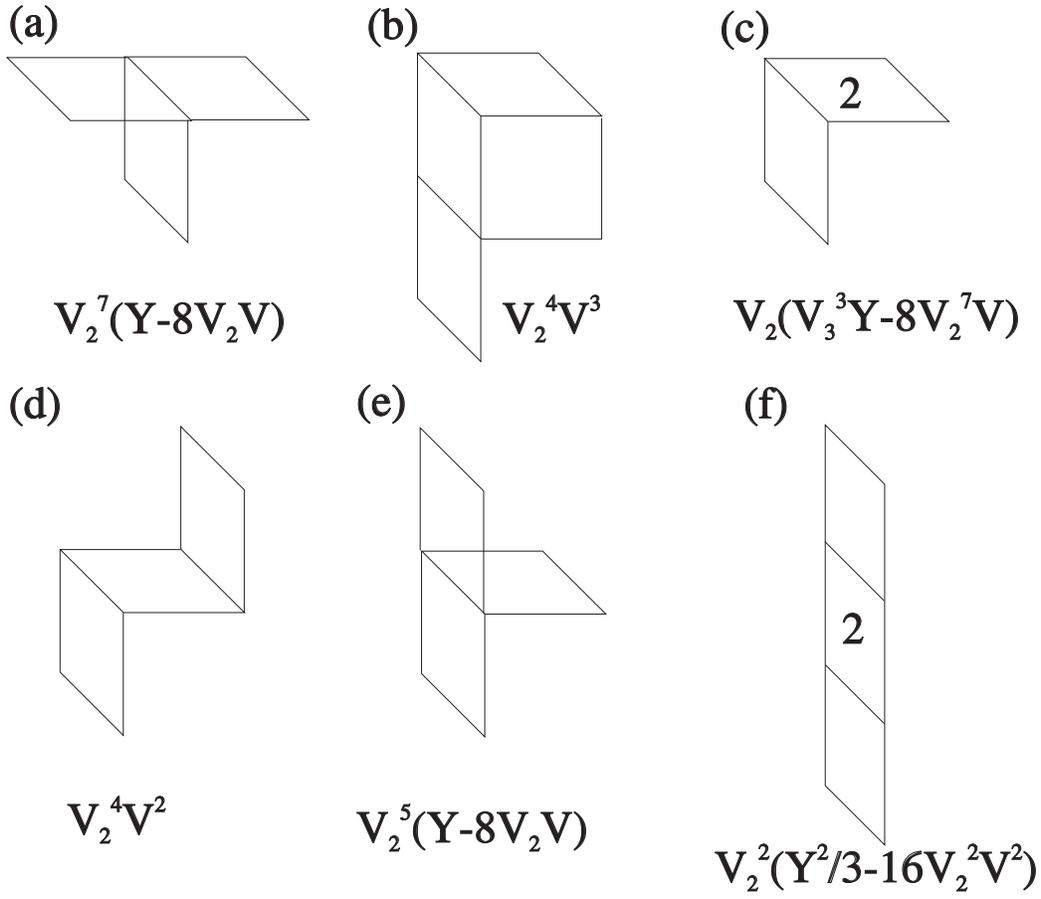}
	\caption{Connected expectations of some of the basic diagrams
	used in calculating the correlation $C(t)$.  The symbols used are
	defined in the text.}
	\label{fig-exdiags}
\end{figure}

\begin{figure}
	\epsffile{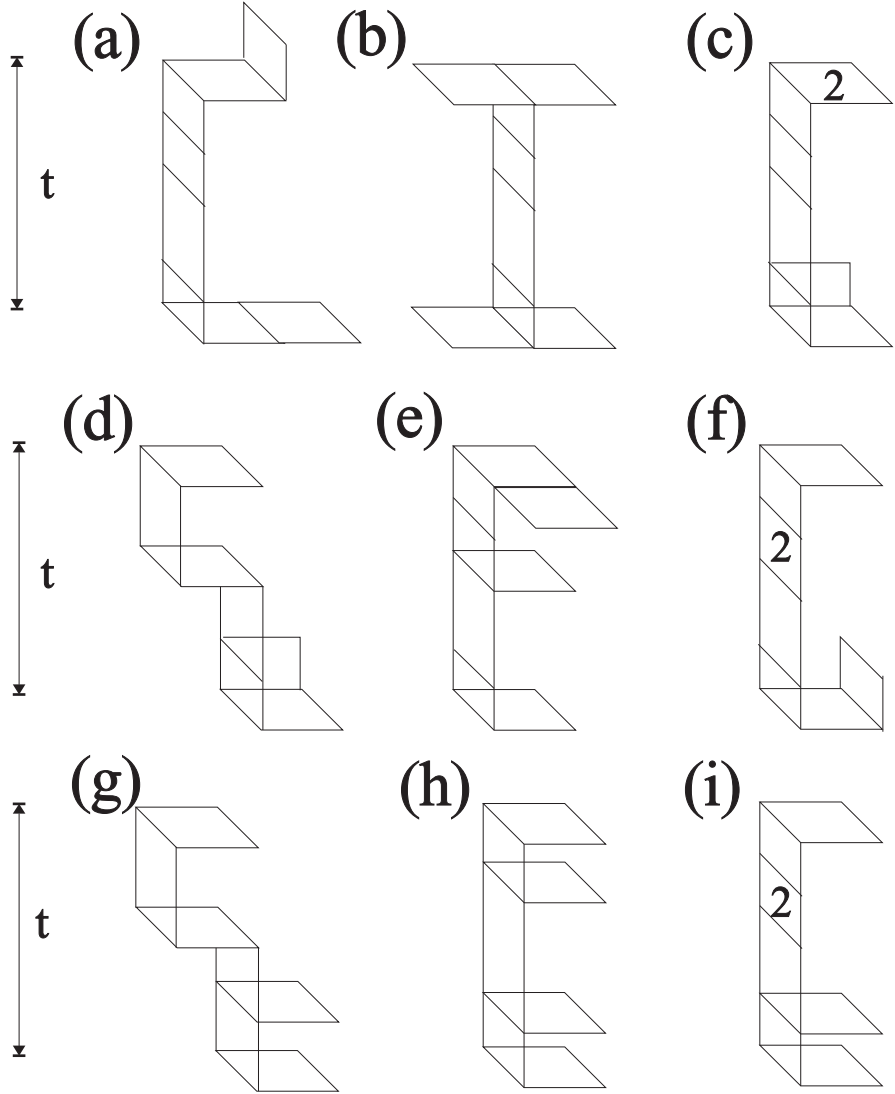}
	\caption{Examples of non-vanishing diagrams at
	$O(\delta^{t+2})$ in the delta expansion for the $SU(2)$ mass
	gap.  The vertical direction represents time.  Diagrams (a)-(c)
        have multiplicity independent of $t$, (d)-(f) have multiplicity
        $\propto t$, and (g)-(i) have multiplicity $\propto t^2$.}
\label{fig-diags2}
\end{figure}

\begin{figure}
	\epsffile{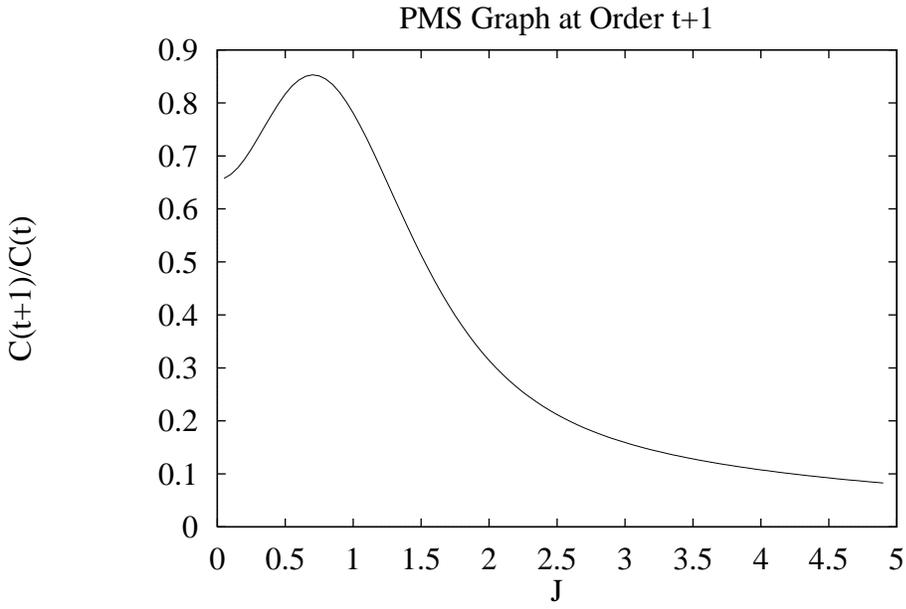}
	\caption{$O(\delta^{t+1})$ PMS graph ($\beta=2.62$).}
	\label{fig-pms1}
\end{figure}

\begin{figure}
	\epsffile{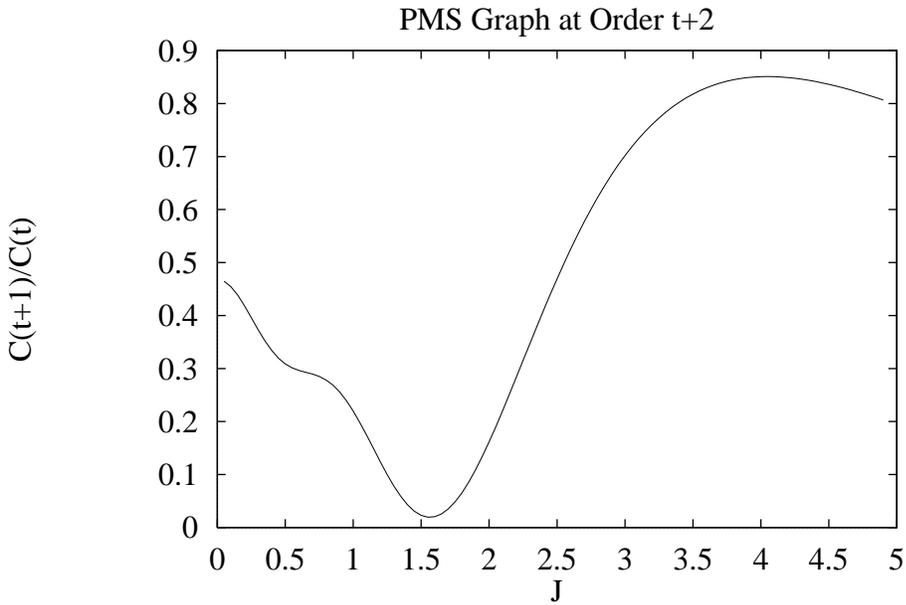}
	\caption{$O(\delta^{t+2})$ PMS graph ($\beta=2.64$).}
	\label{fig-pms2}
\end{figure}

\begin{figure}
	\epsffile{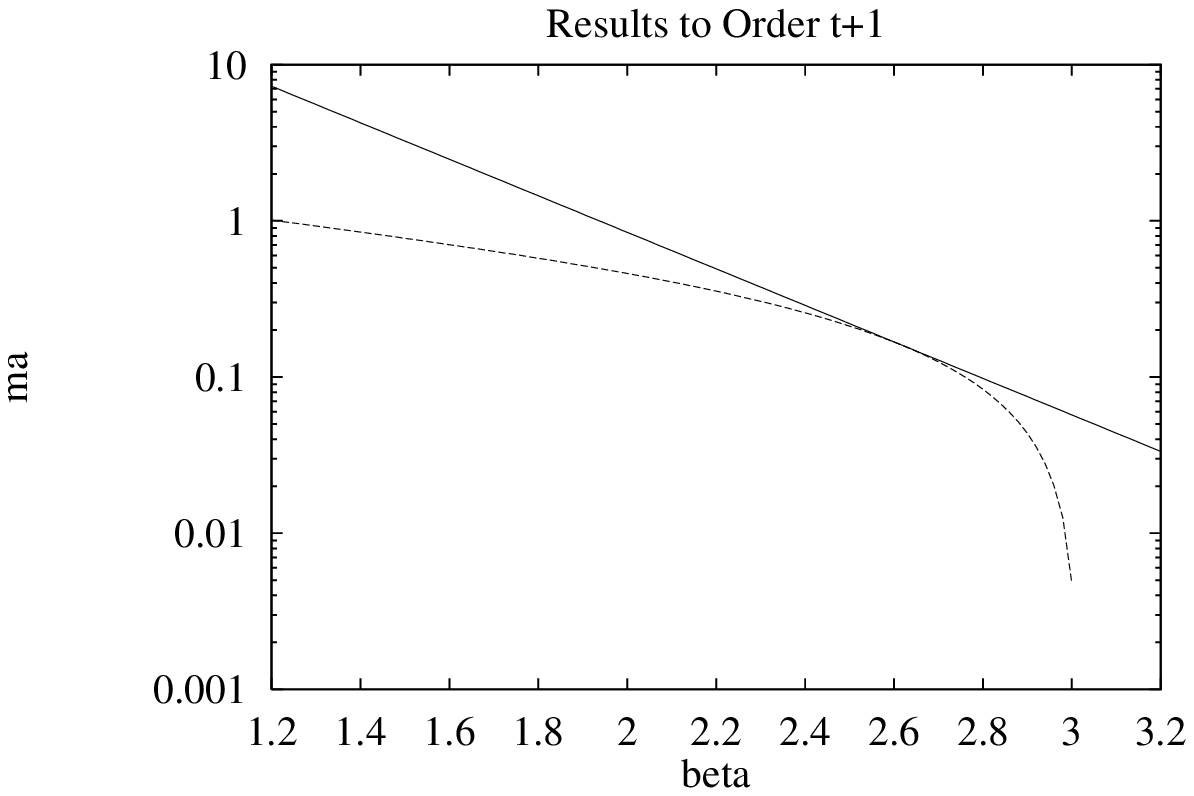}
	\caption{$O(\delta^{t+1})$ results for $ma$ (dotted line), 
        compared with weak coupling renormalization group formula 
	for $\Lambda_L a$ (solid line), with $C_m = 184$.}
	\label{fig-graph1}
\end{figure}

\begin{figure}
	\epsffile{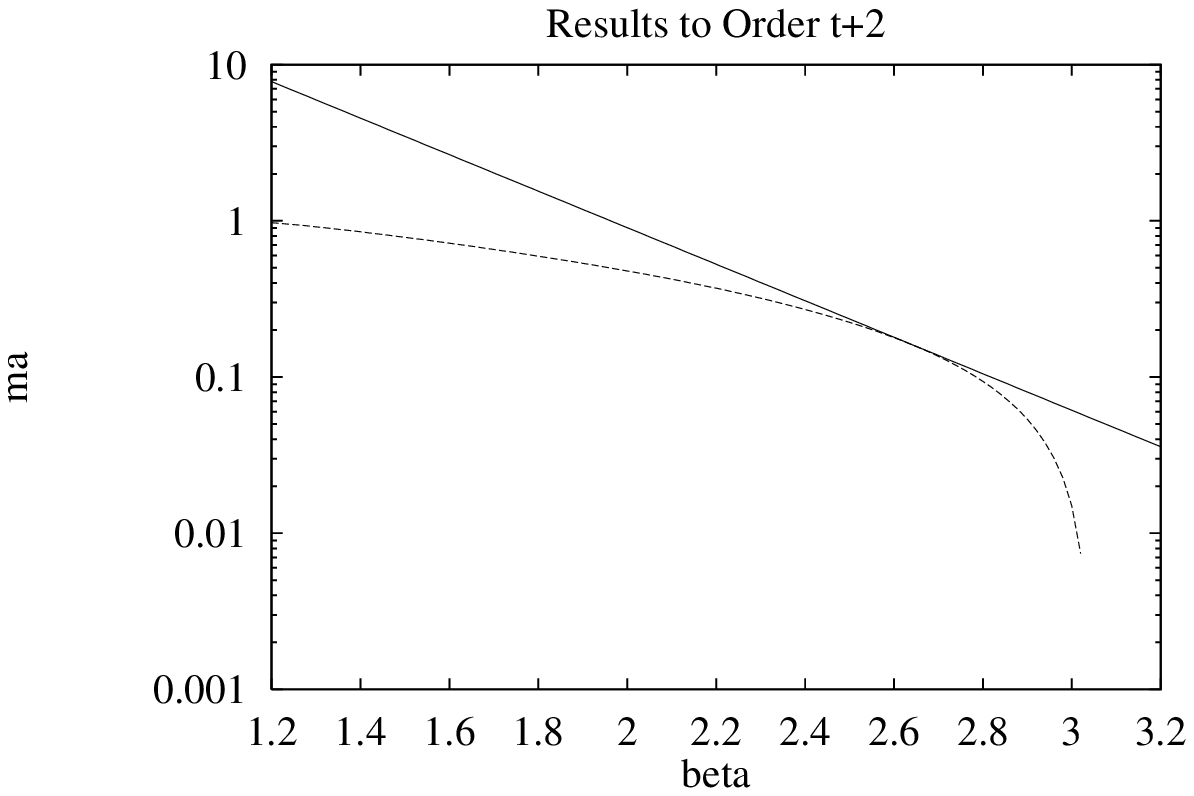}
	\caption{$O(\delta^{t+2})$ results for $ma$ (dotted line), 
	compared with weak coupling renormalization group formula 
	for $\Lambda_L a$ (solid line), with $C_m = 197$.}
	\label{fig-graph2}
\end{figure}

\end{document}